\documentclass[prl,twocolumn,superscriptaddress,showpacs]{revtex4-1}
\usepackage{graphicx,amsmath,amssymb,bm,nicefrac}

\newcommand{\be}{\begin{equation}}	
\newcommand{\ee}{\end{equation}}
\newcommand{\vlowk}{V_{{\rm low}\,k}}

\newcommand{\fmi}{\, \text{fm}^{-1}}
\newcommand{\mev}{\, \text{MeV}}
\newcommand{\kev}{\, \text{keV}}

\newcommand{\sdfp}{s d f_{\nicefrac{7}{2}} p_{\nicefrac{3}{2}}}
\newcommand{\pfg}{p f g_{\nicefrac{9}{2}}}
\newcommand{\du}{d_{\nicefrac{5}{2}}}
\newcommand{\dl}{d_{\nicefrac{3}{2}}}
\newcommand{\s}{s_{\nicefrac{1}{2}}}
\newcommand{\fs}{f_{\nicefrac{7}{2}}}
\newcommand{\pt}{p_{\nicefrac{3}{2}}}
\newcommand{\ff}{f_{\nicefrac{5}{2}}}
\newcommand{\po}{p_{\nicefrac{1}{2}}}
\newcommand{\gn}{g_{\nicefrac{9}{2}}}

\begin{document}

\title{Three-body forces and proton-rich nuclei}

\author{J.\ D.\ Holt}
\affiliation{Department of Physics and Astronomy, University of 
Tennessee, Knoxville, TN 37996, USA}
\affiliation{Physics Division, Oak Ridge National Laboratory, P.O. 
Box 2008, Oak Ridge, TN 37831, USA}
\author{J.\ Men\'{e}ndez}
\affiliation{Institut f\"ur Kernphysik, Technische Universit\"at
Darmstadt, 64289 Darmstadt, Germany}
\affiliation{ExtreMe Matter Institute EMMI, GSI Helmholtzzentrum f\"ur
Schwerionenforschung GmbH, 64291 Darmstadt, Germany}
\author{A.\ Schwenk}
\affiliation{ExtreMe Matter Institute EMMI, GSI Helmholtzzentrum f\"ur
Schwerionenforschung GmbH, 64291 Darmstadt, Germany}
\affiliation{Institut f\"ur Kernphysik, Technische Universit\"at
Darmstadt, 64289 Darmstadt, Germany}

\begin{abstract}
We present the first study of three-nucleon (3N) forces for
proton-rich nuclei along the $N=8$ and $N=20$ isotones. Our results
for the ground-state energies and proton separation energies are in
very good agreement with experiment where available, and with the
empirical isobaric multiplet mass equation. We predict the spectra
for all $N=8$ and $N=20$ isotones to the proton dripline, which
agree well with experiment for $^{18}$Ne, $^{19}$Na, $^{20}$Mg and
$^{42}$Ti. In all other cases, we provide first predictions based on
nuclear forces. Our results are also very promising for studying
isospin symmetry breaking in medium-mass nuclei based on chiral
effective field theory.
\end{abstract}


\maketitle

Exotic nuclei with extreme ratios of neutrons to protons can become
increasingly sensitive to new aspects of nuclear forces. This has been
shown in shell model studies with three-body forces for the
neutron-rich oxygen~\cite{Oxygen,Oxygen2} and calcium~\cite{Calcium}
isotopes, which present key regions for exploring the evolution to the
neutron dripline and for understanding the formation of shell
structure. Calculations with three-nucleon forces predicted an increase in binding of
the neutron-rich $^{51,52}$Ca isotopes compared to existing
experimental values, which was recently confirmed by high-precision
Penning-trap mass measurements~\cite{Gallant}. The pivotal role of
3N forces has also been highlighted in large-space coupled-cluster
calculations~\cite{CC,CCRoth}.

Proton-rich nuclei provide complementary insights to strong
interactions, exhibit new forms of radioactivity, and are key for
nucleosynthesis processes in astrophysics, such as the
rapid-proton-capture process that powers X-ray
bursts~\cite{Blank,Pfuetzner}. Although the proton dripline is
significantly better constrained experimentally than the neutron
dripline, nuclear forces remain unexplored in medium-mass proton-rich
nuclei. Because the proton dripline is closer to the line of
stability, it has also been mapped out empirically by calculating the
energies of proton-rich systems from known neutron-rich nuclei using
the isobaric multiplet mass equation (IMME)~\cite{IMME,Ormand} or
Coulomb displacement energies~\cite{Brown}. This suggests that
proton-rich nuclei provide an important testing ground for nuclear
forces including known Coulomb and isospin-symmetry-breaking effects.

\begin{figure}[t]
\begin{center}
\includegraphics[scale=0.62,clip=]{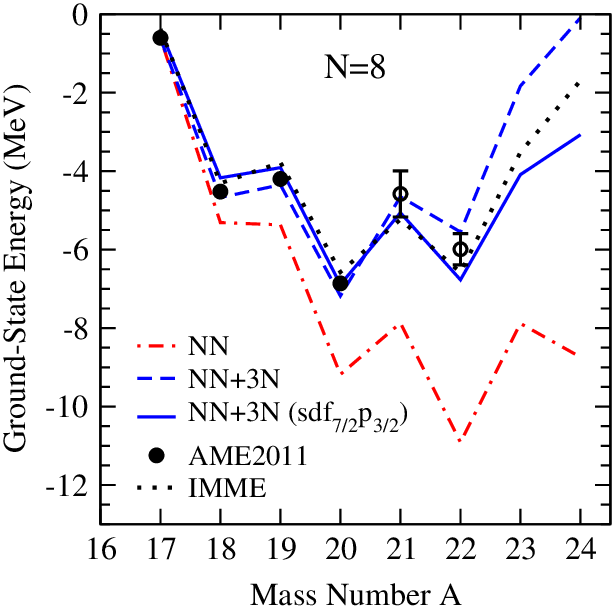}
\end{center}
\caption{Ground-state energies of $N=8$ isotones relative to~$^{16}$O.  
Experimental energies (AME2011~\cite{AME2011} with extrapolations as
open circles) and IMME values are shown. We compare NN-only results
in the $sd$-shell to calculations based on NN+3N forces in both 
$sd$ and $\sdfp$ valence spaces with the consistently calculated
SPEs of Table~\ref{spetab}.\label{N8_gs}}
\end{figure}

\begin{figure}[t]
\begin{center}
\includegraphics[scale=0.36,clip=]{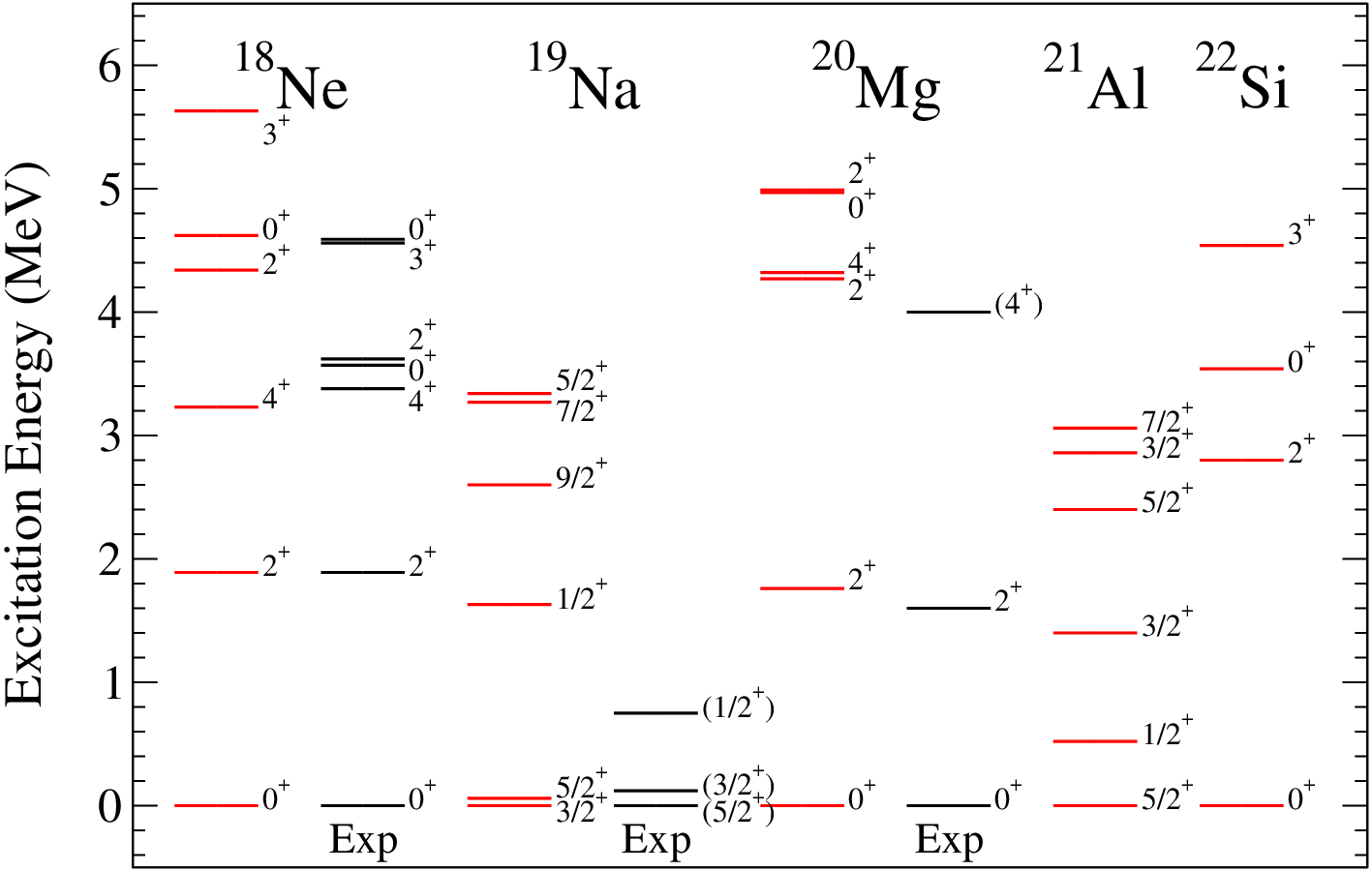}
\end{center}
\caption{Excitation energies of $N=8$ isotones calculated with 
NN+3N forces in the $\sdfp$ valence space, compared with experimental
data~\cite{ENSDF,Angulo19Na,Mukha19Na,Gade20Mg,Mukha} where 
available.\label{N8_spec}}
\vspace*{-4mm}
\end{figure}

\begin{figure}[t]
\begin{center}
\includegraphics[scale=0.62,clip=]{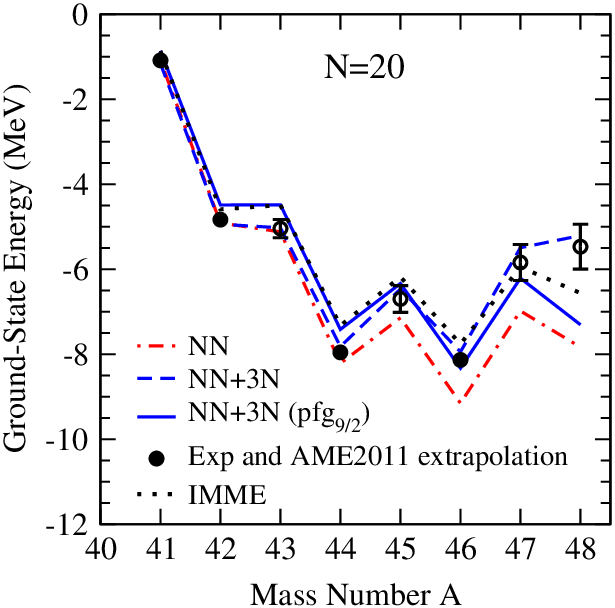}
\end{center}
\caption{Ground-state energies of $N=20$ isotones relative 
to~$^{40}$Ca. Experimental energies~\cite{Dossat} (closed circles)
and AME2011 extrapolations~\cite{AME2011} (open circles), as well
as IMME values are shown. We compare NN-only results in the 
$pf$-shell to calculations based on NN+3N forces in both 
$pf$ and $\pfg$ valence spaces with the 
SPEs of Table~\ref{spetab}.\label{N20_gs}}
\end{figure}

\begin{figure}[t]
\begin{center}
\includegraphics[scale=0.36,clip=]{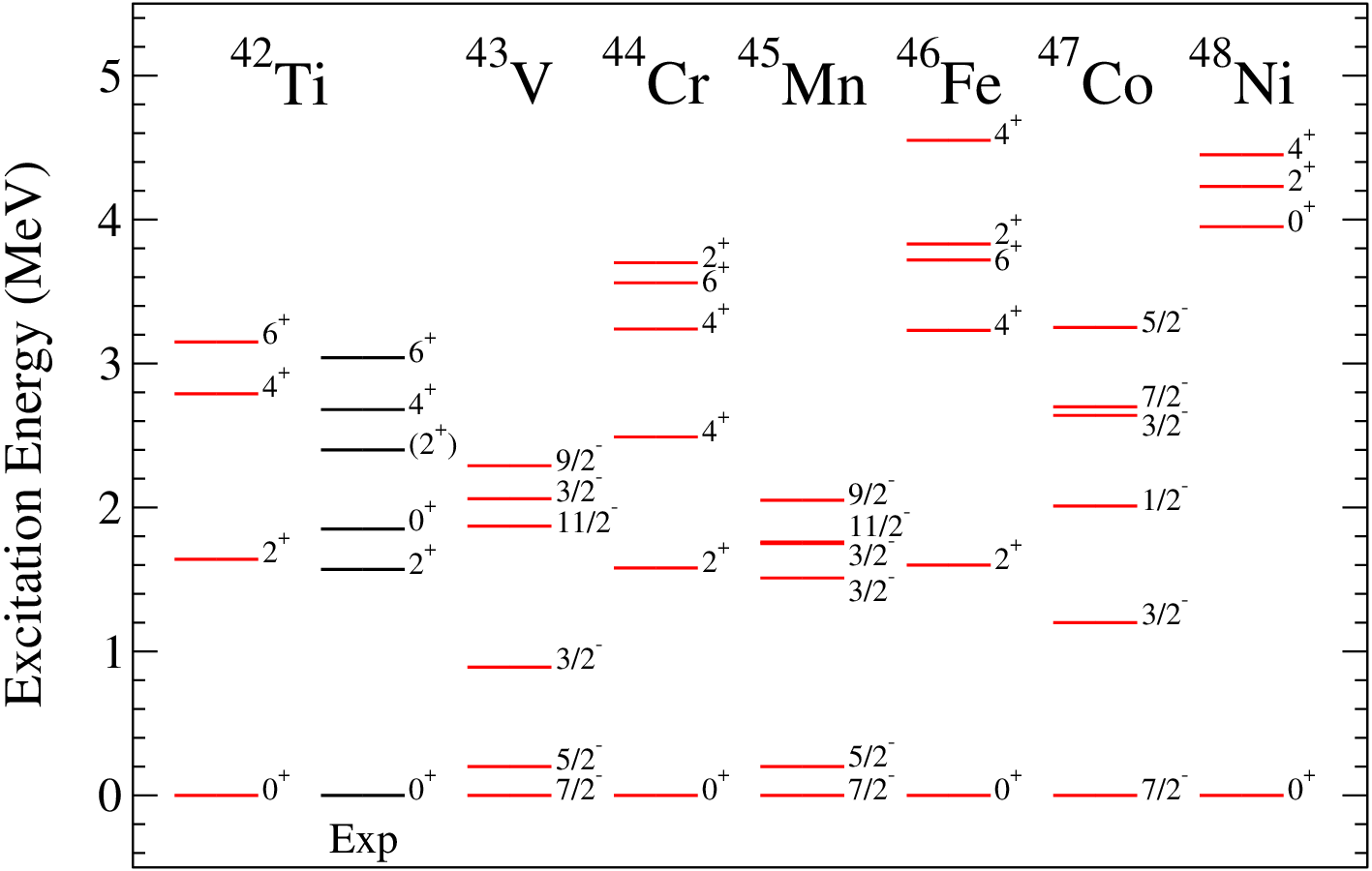}
\end{center}
\caption{Excitation energies of $N=20$ isotones calculated with 
NN+3N forces in the $\pfg$ valence space, compared with experiment
available for $^{42}$Ti only~\cite{ENSDF}.\label{N20_spec}}
\vspace*{-2.5mm}
\end{figure}

In this Letter, we present the first study of 3N forces for
proton-rich nuclei. The couplings of 3N forces are fit to few-nucleon
systems only, and we provide predictions for the ground-state energies
(Figs.~\ref{N8_gs} and~\ref{N20_gs}) and spectra (Figs.~\ref{N8_spec}
and~\ref{N20_spec}) along the chains of $N=8$ and $N=20$ isotones to
the proton dripline. For the interactions studied here, 3N forces provide
repulsive contributions as protons are added, similar to the
neutron-rich case. This is expected due to the Pauli principle
combined with the leading two-pion-exchange 3N
forces~\cite{Oxygen}. Our results suggest a two-proton-decay candidate
$^{22}$Si, whose $Q$ value is very sensitive to the calculation;
within theoretical uncertainties it could also be loosely bound. For
the $N=20$ isotones, we predict the dripline at $^{46}$Fe and the
two-proton emitter $^{48}$Ni~\cite{Dossat48Ni,Pomorski}. Furthermore,
we find good agreement with experimental spectra of $^{18}$Ne,
$^{19}$Na, $^{20}$Mg and $^{42}$Ti and provide predictions for the
isotones where excited states have not been measured.

We consider a shell model description of the $N=8$ and $N=20$ isotones
and determine the interactions among valence protons, on top of a
$^{16}$O and $^{40}$Ca core, based on nuclear forces from chiral
effective field theory (EFT)~\cite{chiral}. At the NN level, we take
the chiral N$^3$LO potential of Ref.~\cite{N3LO} and evolve to a
low-momentum interaction $\vlowk$ with cutoff $\Lambda = 2.0 \fmi$
using renormalization-group methods, which improve the many-body
convergence~\cite{Vlowk}. Three-nucleon forces are included at the
N$^2$LO level. These consist of the long-range two-pion-exchange part,
as well as one-pion-exchange and short-range contact
terms~\cite{chiral}. The shorter-range 3N couplings $c_D$ and $c_E$
are determined by fits to the $^3$H binding energy and the $^4$He
radius for $\Lambda_{\rm 3N} = \Lambda =2.0 \fmi$~\cite{3Nfit},
without further adjustments in the many-body calculations presented
here. Note that 3N forces depend on the NN interaction used, so that
the contributions from 3N forces differ depending on the cutoff in
chiral NN potentials, and when used with bare chiral interactions
(see, e.g., Ref.~\cite{CCRoth}) versus with renormalization-group-evolved interactions.

Excitations outside the valence space are included to third order in
many-body perturbation theory (MBPT)~\cite{BK,LNP+Gmatrix} in a space of
13 major harmonic-oscillator shells. We have checked that the matrix
elements are converged in terms of intermediate-state excitations. For
the $N=8$ isotones, we consider both the standard $sd$-shell and an
extended $\sdfp$ valence space with $\hbar \omega = 13.53 \mev$, and
for $N=20$, the $pf$ and $\pfg$ spaces with $\hbar \omega = 11.48
\mev$. The extended valence spaces proved important in this framework
for oxygen and calcium isotopes~\cite{Oxygen2,Calcium,Gallant}. In
addition to the NN-force contributions, we include the normal-ordered
(with respect to the core) one- and two-body parts of 3N forces in 5
major shells~\cite{Oxygen2,Gallant}. The normal-ordered parts
dominate over the contributions from residual three-body
interactions~\cite{CC3N,CCRoth}. The latter are expected to be weaker
in normal Fermi systems due to phase-space limitations in the valence
shell compared to the core~\cite{Fermi}.

\begin{table}[t]
\begin{center}
\begin{tabular*}{0.48\textwidth}{ccc|ccc}
\hline\hline
\hspace*{-1.75mm}
orbital \, & \,\, emp & \,\,\, MBPT \, & 
orbital \, & \,\, emp & \,\,\, MBPT \, \\ \hline
$\du$ & $-0.60$ & \, $-0.62$/$-0.41$ & 
$\fs$& $-1.07$ & \, $-1.16$/$-0.86$ \\ 
$\s$  & $-0.10$ & $\,\, 0.82$/$0.95$ & 
$\pt$ & $\,\,\,\, 0.63$ & $\,\, 0.28$/$1.40$ \\ 
$\dl$ & $\,\,\,\,\, 4.40$ & $\,\, 4.30$/$4.57$ & 
$\po$ & $\,\,\,\,\, 2.38$ & $\,\, 2.40$/$3.94$ \\ 
$\fs$ & & $\,\,\,\,\,\,\,\,\,\,\,\,\,\,\, 9.73$ & $\ff$ &$\,\,\,\,\, 5.00$ & 
$\,\, 4.91$/$5.36$ \\ 
$\pt$ & & $\,\,\,\,\,\,\,\,\,\,\,\,\, 12.64$ & 
$\gn$ & & $\,\,\,\,\,\,\,\,\,\,\,\,\,\,\, 6.40$ \\[0.5mm]
\hline\hline
\end{tabular*}
\vspace*{-2.5mm}
\caption{Empirical (emp) and calculated (MBPT in the standard/extended valence 
spaces) SPEs in MeV.\label{spetab}}
\vspace*{-4mm}
\end{center}
\end{table}

For the valence proton single-particle energies (SPEs) in $^{17}$F and
$^{41}$Sc, we solve the Dyson equation self-consistently including the
contributions from NN and 3N forces in the same spaces and to the same
order in MBPT. Our calculated SPEs are given in Table~\ref{spetab}, in
comparison with empirical SPEs taken from the experimental spectra of
$^{17}$F and $^{41}$Sc. The MBPT SPEs are similar to the empirical
values, but the $\s$ and $\pt$ orbitals are at higher energy in both
spaces. This finding is similar to the calculated neutron SPEs in
oxygen and calcium isotopes~\cite{Oxygen2,Calcium}, which are more
bound and differ due to Coulomb and isospin-symmetry-breaking
interactions.

All calculations based on NN forces are performed with the empirical
SPEs in the standard $sd$- and $pf$-shells (NN forces only lead to
poor SPEs), while those involving 3N forces use MBPT SPEs in both
standard and extended valence spaces. In this work, we focus on 3N
forces, whose contributions are of the order of a few MeV, but an
explicit inclusion of the continuum naturally becomes important for
weakly bound states and can lead to very interesting contributions,
typically of several hundred keV~\cite{CC,open}. Therefore, we only
show spectra to $^{22}$Si and to the two-proton emitter
$^{48}$Ni. Note that in the case of weakly bound or unbound states,
additional attractive contributions from the continuum are
expected~\cite{open}.

We first consider the ground-state energies of the $N=8$ isotones from
$^{18}$Ne to $^{24}$S, which we compare with experiment when
available. As data is limited, we also employ the IMME~\cite{IMME}.
This relates the energies in an isospin multiplet (of states with the
same quantum numbers in different isobars $A$) by a quadratic
dependence in isospin projection $T_z=(Z-N)/2$,
\begin{equation}
E(A,T,T_z)=a(A,T)+b(A,T) \, T_z+c(A,T) \, T_z^2 \,.
\end{equation}
The energies of proton-rich nuclei can thus be obtained from their
$-T_z$ isobaric analogues via $E(A,T,T_z)=E(A,T,-T_z)+2b(A,T)T_z$,
using a standard fit of the empirical $b$-coefficient~\cite{IMME},
$b=(0.7068A^{2/3}-0.9133) \mev$, with the atomic mass evaluation
(AME2011)~\cite{AME2011} for known neutron-rich nuclei. Moreover, we
include for comparison the extrapolated values of AME2011, although
the IMME is considered to be more accurate.

\begin{table}[t]
\begin{center}
\begin{tabular*}{0.48\textwidth}{c|ccc|ccc}
\hline\hline
& \multicolumn{3}{c|}{$S_p$} & 
\multicolumn{3}{c}{$S_{2p}$} \\
nucleus & exp & \multicolumn{2}{c|}{NN+3N} & 
exp & \multicolumn{2}{c}{NN+3N} \\
$N=8$ & [IMME] & $sd$ & $\sdfp$ \, & [IMME] & $sd$ & $\sdfp$ \\ \hline
$^{18}$Ne & \,\,\,\,\,3.92 & \,\,\,\,\,4.05 & \,\,\,\,\,3.76 
& 4.52 & \,\,\,\,\,4.67 & 4.17 \\ 
$^{19}$Na & $-0.32$ & $-0.32$ & $-0.26$ & 3.60 & \,\,\,\,\,3.73 & 3.50 \\ 
$^{20}$Mg & \,\,\,\,\,2.66 & \,\,\,\,\,2.83 & \,\,\,\,\,2.98 
& 2.34 & \,\,\,\,\,2.51 & 2.72 \\ 
$^{21}$Al & [$-1.34$] & $-2.52$ & $-1.83$ & [1.45] & \,\,\,\,\,0.30 & 1.15 \\ 
$^{22}$Si & \,\,\,\,[1.35] & \,\,\,\,\,0.90 & \,\,\,\,\,1.71
& [0.01] & $-1.63$ & $-0.12$ \\ 
\hline\hline
\end{tabular*}
\vspace*{-2.5mm}
\caption{Experimental and calculated one- and two-proton separation 
energies $S_p$ and $S_{2p}$ (in MeV) of $N=8$ isotones. Where data 
is unavailable, IMME values are given in brackets.\label{N8_Sp}}
\vspace*{-4mm}
\end{center}
\end{table}

Figure~\ref{N8_gs} shows the calculated ground-state energies,
obtained from exact diagonalization in the valence spaces with NN-only
and NN+3N forces, compared with the AME2011 experimental values and
extrapolation, and with the IMME. As expected, the IMME reproduces
well experimental data. It finds $^{22}$Si to be bound, though only by
$10\kev$, with respect to $^{20}$Mg.

In the calculations based on NN forces only, we see a systematic
overbinding throughout the isotone chain, which becomes more
pronounced for larger mass number. Three-nucleon forces provide key
repulsive contributions to ground-state energies and good agreement
with experiment is obtained in both valence spaces. The extended-space
predictions become more bound beyond $^{20}$Mg, the last measured
isotone. For both valence spaces, the proton dripline is predicted at
$^{20}$Mg, though $^{22}$Si is unbound with respect to $^{20}$Mg by
only $0.1 \mev$ in the extended space, compared with $1.6 \mev$ in the
$sd$-shell. This makes a prediction of the dripline
difficult, and an experimental measurement of the $^{22}$Si
ground-state energy would present a decisive constraint for 3N
forces. All calculations find a sharp decrease in binding energy past
$^{22}$Si, clearly indicating the dripline has been reached.

A more detailed picture can be developed from the one- and two-proton
separation energies given in Table~\ref{N8_Sp}. $S_p$ and $S_{2p}$ are
key quantities for determining two-proton emission candidates. In
general, we find good agreement between our calculations and the
experimental (and IMME) values. While the $sd$-shell energies are
slightly closer to experiment for lighter isotones, the extended-space
calculations agree best with the IMME beyond $A=20$, approximately the
same point, $^{21}$O, at which the added valence-space orbitals become
important in the oxygen isotopes in this framework~\cite{Oxygen2}.
 
Spectroscopic data in the $N=8$ isotones exists to $^{20}$Mg. In
Fig.~\ref{N8_spec}, we compare the experimental low-lying states in
$^{18}$Ne, $^{19}$Na, and $^{20}$Mg to those calculated with NN+3N
forces in the $\sdfp$ valence space. Calculations with 3N forces in
the $sd$-shell give very similar spectra up to $^{19}$Na, while for
$^{20}$Mg, $^{21}$Al, and $^{22}$Si they are more compressed than in
Fig.~\ref{N8_spec}. In $^{18}$Ne we find good agreement for the first
excited $2^+$ and $4^+$ states. The ground state and first two excited
states in $^{19}$Na have been measured with tentative spin and parity
assignments~\cite{Angulo19Na,Mukha19Na}. The ordering of the first two
states in our calculation disagrees with the tentative assignments,
but the spacing between them is only $0.1 \mev$.  The
$\nicefrac{1}{2}^+$ state is predicted in our calculation close to the
$\nicefrac{1}{2}^+$ in the mirror $^{19}$O, but $0.9 \mev$ above
experiment. This $^{19}$O-$^{19}$Na $\nicefrac{1}{2}^+$ difference is
a clear example of the Thomas-Ehrman
effect~\cite{Angulo19Na,Thomas-Ehrman}.  Since in our $^{19}$Na
calculation the $\s$ orbital is unbound, continuum coupling is
expected to reduce the $\nicefrac{1}{2}^+$ energy. In $^{20}$Mg, only
information on the first excited state has been
published~\cite{Gade20Mg}, but a second excited state has been
measured recently~\cite{Mukha}. While a tentative assignment of $4^+$
was given to this state, we predict two close-lying states ($2^+$ and
$4^+$) at very similar energy. In our predictions for $^{21}$Al and
$^{22}$Si, we note the high $2^+$ state in $^{22}$Si as a possible
indication of a subshell closure analogous to
$^{22}$O~\cite{Oxygen2}. For all cases, the differences of excitation
energies between these proton-rich nuclei and the corresponding mirror
oxygen isotopes are less than $0.8 \mev$.

\begin{table}
\begin{center}
\begin{tabular*}{0.48\textwidth}{c|ccc|ccc}
\hline\hline
& \multicolumn{3}{c|}{$S_p$} & 
\multicolumn{3}{c}{$S_{2p}$} \\
nucleus & \, exp & \multicolumn{2}{c|}{NN+3N} & 
\, exp & \multicolumn{2}{c}{NN+3N} \\
$N=20$ & \, [IMME] & $pf$ & \,\, $\pfg$ \, 
& \, [IMME] & $pf$ & \,\, $\pfg$ \\ \hline
$^{42}$Ti & \,\,\,\,\,\,3.75 & \,\,\,\,\,3.78 & \,\,\,\,\,3.63 
& \,\,\,\,\,4.83 & \,\,\,\,\,4.94 & \,\,\,\,\,4.49 \\ 
$^{43}$V  & [$-0.10$] & \,\,\,\,\,0.09 & \,\,\,\,\, 0.00
& \,\,\,\,\,[3.62] & \,\,\,\,\,3.87 & \,\,\,\,\,3.62 \\ 
$^{44}$Cr & \,\,\,\,\,[2.84] & \,\,\,\,\,2.79 & \,\,\,\,\,2.93
& \,\,\,\,\,3.12 & \,\,\,\,\,2.88 & \,\,\,\,\,2.93 \\ 
$^{45}$Mn & [$-1.15$] & $-1.35$ & $-1.08$ 
& \,\,\,\,\,[1.69] & \,\,\,\,\,1.44 & \,\,\,\,\,1.85 \\ 
$^{46}$Fe & \,\,\,\,\,[1.58] & \,\,\,\,\,1.48 & \,\,\,\,\,1.99 
& \,\,\,\,\,0.18 & \,\,\,\,\,0.12 & \,\,\,\,\,0.91 \\ 
$^{47}$Co & [$-1.81$] & $-2.45$ & $-2.12$ & [$-0.23$] & $-0.97$ & $-0.13$ \\ 
$^{48}$Ni & \,\,\,\,\,[0.61] & $-0.29$ & \,\,\,\,\,1.09
& \,$-1.28(6)$ & $-2.73$ & $-1.02$ \\ 
\hline\hline
\end{tabular*}
\vspace{-2.5mm}
\caption{Experimental and calculated one- and two-proton separation 
energies $S_p$ and $S_{2p}$ (in MeV) of $N=20$ isotones. Where data 
is unavailable, IMME values are given in brackets. Direct measurements
of $S_{2p}$ in $^{48}$Ni are from Refs.~\cite{Dossat48Ni,Pomorski}.
\label{N20_Sp}}
\vspace*{-4mm}
\end{center}
\end{table}

Next, we show in Fig.~\ref{N20_gs} the ground-state energies of the
$N=20$ isotones from $^{42}$Ti to $^{48}$Ni, where the IMME also
reproduces well the limited experimental
data~\cite{Dossat}. Calculations with NN forces already lead to a
reasonable description of experiment, with energies only modestly
overbound (within $1 \mev$) beyond $^{45}$Mn. When 3N forces are
included, the additional repulsion systematically improves the
agreement with data. The extended-space calculations agree very well
with the IMME throughout the isotone chain, while the $pf$-shell
results deviate for $^{47}$Co and $^{48}$Ni. In all calculations the
proton dripline is robustly predicted at $^{46}$Fe.

The one- and two-proton separation energies are given in
Table~\ref{N20_Sp}. The experimental and IMME values generally fall
within the NN+3N calculations in the $pf$ and $\pfg$ valence
spaces. The difference in $S_p$ and $S_{2p}$ between the two
calculations only becomes larger than $0.7 \mev$ for $^{46}$Fe,
$^{47}$Co and $^{48}$Ni. This indicates that, in our framework, the
$\gn$ orbital becomes relevant around $A=47$ and provides extra
binding, similar to the calcium
isotopes~\cite{Calcium,Gallant}. Indeed, our $\pfg$ result for
$S_{2p}$ of $^{48}$Ni is only $0.3 \mev$ larger than recent
experiment~\cite{Dossat48Ni,Pomorski}.

Spectroscopic data is only available for $^{42}$Ti in the $N=20$
isotones. We show the predicted spectra based on NN+3N calculations in
the $\pfg$ valence space in comparison with experiment in
Fig.~\ref{N20_spec}. The energies of the first $2^+$, $4^+$ and $6^+$
are well reproduced.  There are two observed states between $2^+_1$
and $4^+_1$ that do not appear in our calculation. We attribute these
to neutron (4p2h) excitations, expected around
$^{40}$Ca~\cite{GeraceGreen}. For the remaining isotones, we show our
predictions for the energies of the first five excited states below $5
\mev$. Similar to $^{22}$Si, we note the high energy of the $2^+$
state in $^{48}$Ni as a tentative closed subshell signature. The
excitation energy difference with respect to the mirror calcium
isotopes is smaller than $0.3 \mev$, in agreement with the
experimental knowledge in this region~\cite{med}. The calculated
spectra in the $pf$-shell are similar, though modestly compressed, up
to $^{44}$Cr, and more compressed beyond.

In summary, we have presented the first study of 3N forces in
proton-rich medium-mass nuclei. Our results for ground- and
excited-state energies are in very good agreement with experiment,
including the prediction of a recently discovered state in
$^{20}$Mg~\cite{Mukha}. A future measurement of the ground-state
energy of $^{22}$Si will provide an important constraint for 3N
forces. We make predictions for the unexplored spectra of the $N=8$
and $N=20$ isotones. Our extended-space calculations for the
ground-state energies are of similar quality as empirical IMME
predictions, which is very promising for studying isospin symmetry
breaking in medium-mass nuclei based on chiral EFT interactions. Our
work presents a bridge to future studies, based on nuclear forces, of
exotic nuclei with proton and neutron valence degrees of freedom.

We thank M.\ Pf\"{u}tzner, B.\ Blank, I.\ Mukha and A.\ Poves for
helpful discussions, and the TU Darmstadt for hospitality. This work
was supported by the US DOE Grant DE-FC02-07ER41457 (UNEDF SciDAC
Collaboration), DE-FG02-96ER40963 (UT), the Helmholtz Alliance Program
of the Helmholtz Association, contract HA216/EMMI ``Extremes of
Density and Temperature: Cosmic Matter in the Laboratory'', the BMBF
under Contract No.~06DA70471, and the DFG Grant SFB 634. Part of the
calculations were performed on Kraken at the National Institute for
Computational Sciences.

\end{document}